\documentstyle[aps,epsf,array]{revtex}
\def\eqnn#1{(\ref{#1})}
\def\plb{{ \sl Phys. Lett. }}

\def\jsp{{\sl  J. Stat. Phys.}}
\def\figno#1{Fig.~\ref{fig:#1}}
\def\cum#1{\langle\langle#1\rangle\rangle}

\begin{document}
\title{On the Violations of Local Equilibrium and Linear Response}
\author{Kenichiro Aoki\cite{ken-email}
  and Dimitri Kusnezov\cite{dimitri-email}  }
\address{$^a$Dept. of Physics, Keio University, {\it
    4---1---1} Hiyoshi, Kouhoku--ku, Yokohama 223--8521, Japan\\
  $^b$Center for Theoretical Physics, Sloane Physics Lab, Yale
  University, New Haven, CT\ 06520-8120} \date{\today }
\maketitle
\begin{abstract}
  We study how local equilibrium, and linear response
  predictions of transport 
  coefficients are violated as systems move far from equilibrium.
  This is done by studying heat flow in classical lattice models
  with and without bulk transport behavior, in 1--3 dimensions. We 
  see that linear response and local equilibrium assumptions
  break down at the same rate. The equation of state is also
  found to develop non-local corrections in the steady state. We
  quantify the breakdown through the analysis of both
  microscopic and macroscopic observables, which are found to
  display non-trivial size dependence.
\end{abstract}
\vspace{3mm}
In studies of non-equilibrium systems, local equilibrium is 
an assumption which is essential to allowing
the use of statistical mechanics and equilibrium or
non--equilibrium thermodynamics\cite{degroot}. 
Without local equilibrium, even the definition of temperature is
not unique\cite{non-eq-Temp}, and it becomes unclear how to define simple
transport processes. Local equilibrium is typically justified
through conditions 
argued to be necessary or sufficient. This might involve
checking for Onsager reciprocity\cite{onsager}, enforcing upper limits on
local fluctuations of temperature (density,...)\cite{fluct},
verifying that the 
equation of state holds locally\cite{eos}, and so
forth\cite{hk}. However, these conditions 
and inequalities do not offer any quantitative guidance into how
local equilibrium breaks down. Further, linear response has not been
tested in conjunction with the breakdown of local equilibrium,
which is an important consideration if one questions whether or not
higher order corrections to Fourier's Law are consistent. 
Previous studies have observed the breaking of local
equilibrium\cite{local-eq,dhar} as well as deviations from
linear response\cite{takesue,ak-long}. However few
such cases are known, and the
quantitative behavior of physical observables when
local equilibrium is broken has not been studied previously. 
In \cite{dhar}, the breakdown of local equilibrium was observed in the
X--Y model and the Lorentz gas under thermal
gradients, which was  attributed 
to the infinite number of local conservation laws in
the dynamics.  Our results for non--integrable models will show that
the integrability of the systems are not necessary for the
deviations from local equilibrium to occur.
In this letter we establish a quantitative guide to the {\sl rate} at
which concepts like local equilibrium and linear response
become  violated in systems which are subject to thermal
gradients. We study lattice models in $d=1-3$ spatial dimensions,
including the Fermi-Pasta-Ulam $\beta$ model, which does not
have a bulk transport limit in $d=1$. 
We also test how the equation of state is modified and its
relation to expectations from irreversible thermodynamics.

How should  physical observables behave away from local
equilibrium? A natural idea is that a
physical observable  $\cal A$  will deviate from its value in local
equilibrium as we move further away from equilibrium.  In our
case, a temperature gradient $\nabla T$ provides a natural
measure of how far we are from equilibrium.  Since the intrinsic
physical properties should not depend on which side of the
box is at a higher temperature, the deviation $\delta {\cal A}$ from
its local equilibrium value is expected to behave as
\begin{equation}
  \label{o-le}
  \delta_{\cal A}\equiv
  {\delta {\cal A}\over{\cal A}}=
  C_{\cal A}\left(\nabla T\over T\right)^2
  +C'_{\cal A}\left(\nabla T\over T\right)^4
  +\ldots
\end{equation}
While seemingly natural, such an analytic expansion is not
trivial;
in sheared fluids, transport coefficients have been seen to
display non-analytic dependences on the shear rate, which have
not been entirely clarified\cite{shears}.  The coefficients
$C_{\cal A},C'_{\cal A},\ldots$ are in principle dependent on
$T$ and $L$, the size of the lattice in the direction of the
gradient. If the relation is completely local, we expect that
they will be independent of $L$. We shall find that the
situation is more subtle.

We study  two systems which display {\it qualitatively different}
transport behavior. Their Hamiltonians are
\begin{equation}  \label{ham}
  H =\frac{1}{2}\sum_{\bf r}\left[p_{\bf r}^2 +
    \left(\nabla \phi_{\bf r}\right)^2 + V\right].
\end{equation}
where  $V=\beta
(\nabla \phi_{\bf r})^4/2$ for the FPU$-\beta$ model and
$V=\phi_{\bf r}^4/2$ for the $\phi^4$ model. The
$\phi^4$ model has a well defined bulk limit for the thermal
conductivity in $d=1-3$\cite{ak-long} whereas the FPU model has $L$
{\it dependent} thermal conductivity in
$d=1$\cite{fpu-later,ak-fpu}. 
We will use $\beta=1$ without loss of generality.
Here ${\bf r}$ runs over all
sites in the lattice ($x=1,...,L$, $y,z=1,...,N_\perp$), 
and the lattice derivative has components
$\nabla_k\phi_{\bf r}\equiv \phi_{\bf r + e_k}-\phi_{\bf r}$
(${\bf e_k}$ is the unit lattice vector in the $k$-th
direction). For $d=2,3$ we take $N_\perp=3-20$ sites in the
transverse directions with periodic boundary conditions. 
We thermostat the endpoints $L=0,N+1$ dynamically at
temperatures $T_1$ and $T_2$, as discussed
in \cite{ak-fpu,ak-long}. 

\begin{figure}[htbp]
  \begin{center}
    \leavevmode    \epsfxsize=8cm\epsfbox{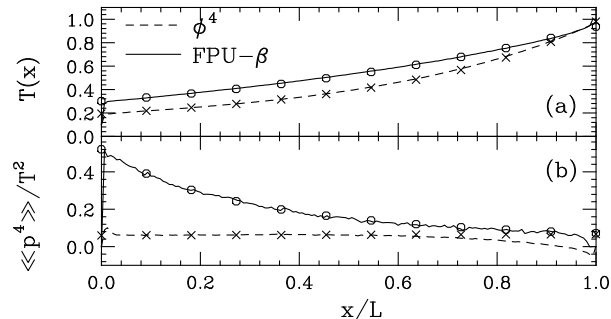}
    \caption{(a) Temperature profiles for the $\phi^4$ theory
    and the FPU model and the corresponding normalized cumulants 
    (b), for $L=162$, with the boundary temperatures
        $(0.1,1.0)$. Symbols ($\circ,\times$) are the analytic
    predictions using 
    $T(x)$ in (a) and  Eq.~\eqnn{o-le1} in (b). The different $\gamma$
    between the two models in this temperature range causes a
    different relative sign between $a$ and $b$ in  Eq.~\eqnn{o-le1}
    accounting for the different shapes in (b).} 
    \label{fig:profs}
  \end{center}
\end{figure}

Once we determine $C_{\cal A}$ for an observable ${\cal A}$, it is
possible to compute its non-equilibrium spatial distribution
function.  When $\kappa(T)$ behaves as a
power law in the temperature range of interest,
denoted $\kappa=cT^{-\gamma}$, the temperature profile  is
$T(x)=T_1(1-(1-(T_2/T_1)^{1-\gamma})x/L)^{1/(1-\gamma)}$\cite{ak-long}.
Such a power law behavior for $\kappa(T)$ has been shown to hold for the
$\phi^4$ model in $d=1-3$  and also in most temperature regions
in the $d=1$ FPU $\beta$ model, including the region we work
with here \cite{ak-long,ak-fpu}.
The agreement for the predicted profile is shown in
\figno{profs}(a). Using Fourier's law ($J=-\kappa\nabla T$),
Eq.~\eqnn{o-le} and  $T(x)$, we derive to leading order:
\begin{equation}
  \label{o-le1}
  {\delta {\cal A}\over{\cal A}}=
  C_{\cal A}\left(J\over \kappa T\right)^2
  =C_{\cal A}\left(\frac{1}{a+bx}\right)^2
\end{equation}
where $a=T_1^{1-\gamma}c/J$, $b=\gamma-1$, and $J$ is the heat
flow.  So knowing $C_{\cal A}$, we can also predict the spatial
variation of the non-equilibrium observable ${\cal A}$. This
bring to light an interesting relation to coarse graining.
Coarse graining in $x$ over regions of 
length $\ell$, with $L>\ell >\lambda$, where $\lambda$ is 
the mean free path of the
excitations, will provide no significant improvement towards
recovering local equilibrium, since (up to the sign of $C_{\cal A}$), the
functional form of \eqnn{o-le1} is positive definite. (This will be
evident in \figno{profs}(b).)

One might also be inclined to expand \eqnn{o-le} in  powers of
$(\nabla^n T)/T$. 
However, in the region where the temperature
dependence of the thermal conductivity  can be described by a
power law, $\kappa=cT^{-\gamma}$, one can show, using Fourier's
law, that $\nabla^n T/T= a_n\times\left( \nabla T/T\right)^n$,
where $a_n$ is a temperature independent constant.  Strictly
speaking, Fourier's law holds only close to local equilibrium,
but as we shall see later, the deviations from it is of order
$(\nabla T/T)^2$ so that the difference is a higher order
correction in the expansion in $(\nabla T/T)$. 

\noindent\underline{Local Equilibrium:}
With the local temperature given by $T_k=\langle
p_k^2\rangle$, a natural measure for the deviations from
local equilibrium is the deviation of the momentum
distribution from the Maxwellian distribution. The cumulants 
$\cum{p_k^4} = \langle p_k^4\rangle -
  3\langle p_k^2\rangle^2$, 
  $\cum{p_k^6} = \langle p_k^6\rangle -
  15 \langle p_k^4\rangle   \langle p_k^2\rangle 
  + 30 \langle p_k^2\rangle ^3$, and so on,
normalized by the local
temperature,  $\cum{p^n}/T^{n/2}$, provide a quantitative
measure on  how far we are from local equilibrium. In local
equilibrium, $\cum{p^n}=0$ ($n>2$).
Consider first how systems typically behave under thermal
gradients.  In \figno{profs}, the
local temperature and  $\cum{p_x^4}/T_x^2$ are plotted against
the position in both $\phi^4$ and FPU systems.
When the
system is not too far from equilibrium, the temperature profile
can be understood by {\it locally} applying Fourier's law
$J=-\kappa(T) \nabla T$, where the heat flow $J$ is 
constant
\cite{ak-long,ak-jumps,tprofs}. Since $\kappa$  depends on $T$, 
$\nabla T$ will also depend on $T$ and
hence the temperature profile becomes curved for increasing
boundary temperature differences.  We note that the
temperature profile for the $\phi^4$ theory is visibly more
curved under the same boundary conditions, even though the FPU
is further from local equilibrium (\figno{profs}(b)).

In \figno{profs}(b), we see that the $\cum
{p_k^4}$ cumulants are non--zero inside the system, so that we
are no longer in local equilibrium. 
Contrary to naive intuition, the steepest gradient
does {\it not } lead to the system being furthest from local
equilibrium.  In fact, the converse is true --- the
system is furthest from equilibrium in the flattest
region.

\begin{figure}[htbp]
  \begin{center}
    \leavevmode\epsfxsize=8.3cm\epsfbox{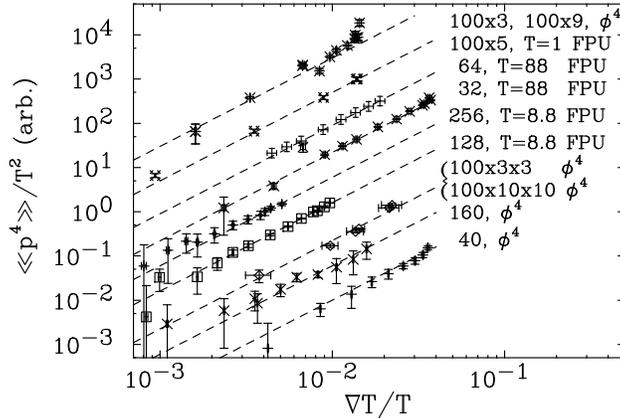}
    \caption{
      $\cum{p^4}/T^2$ away from
      equilibrium for the $\phi^4$ theory ($d=1,2,3,\ T=1$) 
      and the FPU model ($d=1,2$) for various  lattice
      sizes on an  arbitrary scale.      
      The  fits of the data to a behavior const.$\times (\nabla
      T/T)^2$ are plotted for each lattice size.}
    \label{fig:cum}
  \end{center}
\end{figure}

In \figno{cum}, we plot the deviations of the 4-th cumulant
of the momentum for the $\phi^4$ theory and FPU model against
$\nabla T/T$, which corresponds to taking ${\cal A}=\langle
p_x^4\rangle$ in \eqnn{o-le}. 
We are interested
in the 
physics away from the boundaries and we shall always measure the
physical quantities locally, well inside the system, although
boundary effects can be readily understood\cite{ak-fpu}.  
We find that $\nabla T/T$ 
provides a good measure 
of how far we are from equilibrium and in both models, the
cumulants behave as 
\begin{equation}
  \label{le-breaking}
  \delta_{\scriptsize LE}=\frac{ \cum{p^4}}{3T^2}= 
  C_{LE} \left(\frac{\nabla T}{T}\right)^2,
\end{equation}
where $C_{LE}^\phi =1.1(8)L^{0.9(2)} \ (T=1)$ and
$C_{LE}^{FPU}=4.3(4)L^{0.99(2)} \ (T=8.8)$.  Similar
investigations at different $T$ yields a weak $T$ dependence for
$C_{LE}$ which is difficult to establish.  These results are
consistent with $d>1$ in both models at the same
temperatures. Using $C_{LE}^\phi$, $C_{LE}^{FPU}$, we can
predict the shape of $\cum{p^4}/T^2$:  In \figno{profs}(b), the
non-equilibrium distribution (\eqnn{o-le1}; symbols) is compared
to simulation results (lines) agreeing nicely.

Here, a relatively simple picture emerges: as we move away from
equilibrium by increasing the difference in the boundary
temperatures, each point in the interior deviates
from local equilibrium in a predictable manner, {\it without}
any threshold.  Away from equilibrium, local equilibrium is an
approximation that is quite good for small gradients since the
deviations from it only vary as $(\nabla T)^2$.
Similar results hold for  higher momentum cumulants.

 The $L$ dependence of $C_{\cal A}$ is quite intriguing.  
Naive argument suggests that
since the gradients and the cumulants are local, the relation
between them would not depend on $L$, at least in the $\phi^4$
theory where there is a bulk limit.  This turns out not to be
the case. In principle, it is possible that the effect we see
will disappear in the large $L$, bulk limit.  However, this
seems implausible since we have excluded the region within the
mean free path from the boundaries in the above results, using
the properties of the model  extracted in \cite{ak-long,ak-fpu}.

\noindent\underline{Linear Response:} 
Let us now investigate the validity of linear response theory.
This has been discussed previously as one of the criteria for
the breakdown of local equilibrium \cite{hk}, even though no
deviation was seen   there. 
A priori, it is not clear if the linear response law can be used
as a criterion for the breakdown of local equilibrium, rather
than as an indication of higher order equilibrium 
corrections\cite{non-lin-response}.  Such
an analysis assumes that even when linear response is
broken, local equilibrium holds sufficiently well so that
one can unambiguously define the temperature inside the system
(or that we adopt a particular definition for $T$). The linear
response prediction of the heat flow $J$ is obtained by
computing $\kappa$ from 
applying Fourier's law locally. This is denoted as $J_{LR}$, 
and agrees with direct measurements of $J$ in the near
equilibrium limit\cite{ak-long}. 

\begin{figure}[htbp]
  \begin{center}
    \leavevmode    \epsfxsize=8.5cm\epsfbox{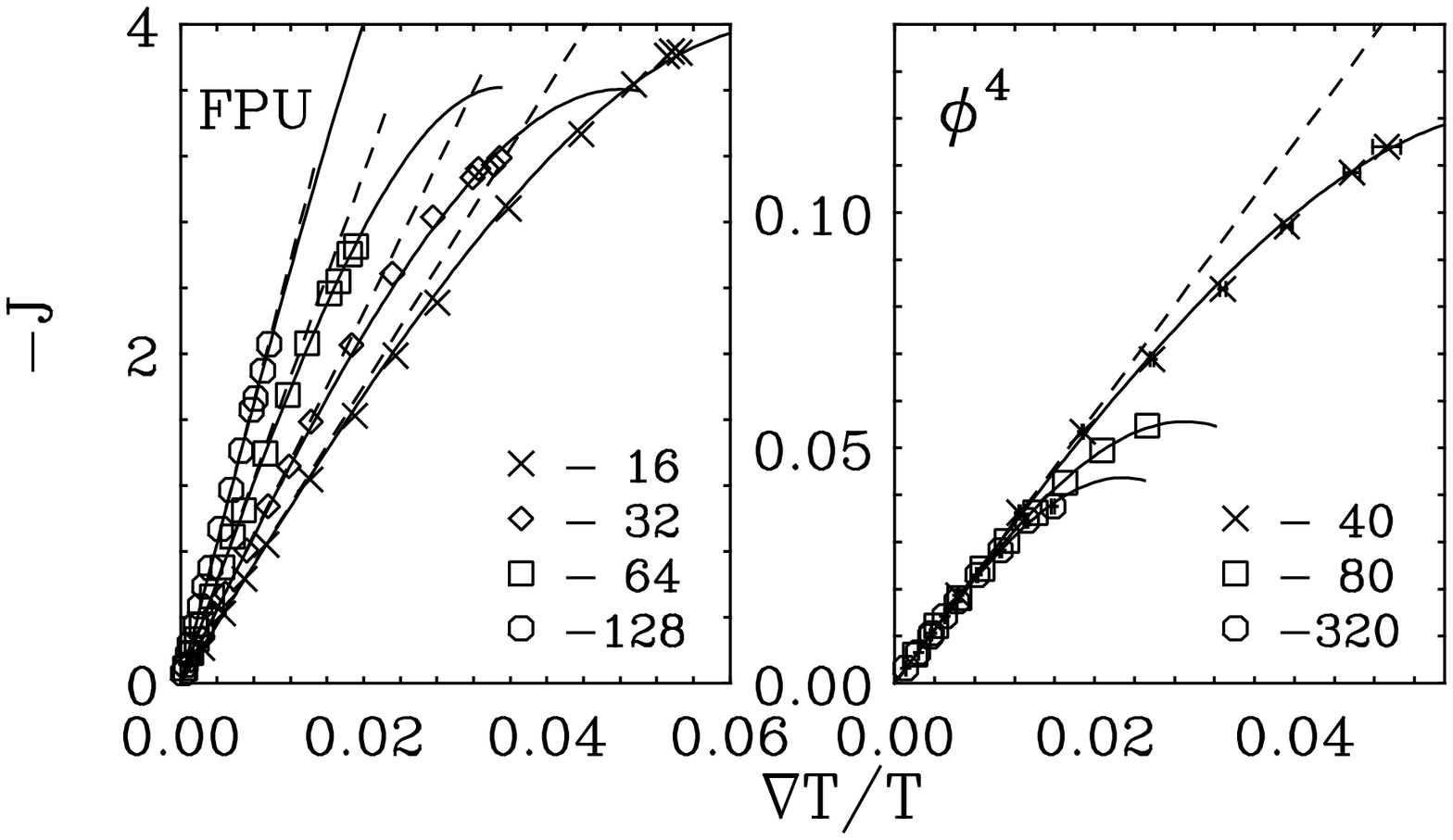}
    \caption{The deviations from the linear response law as a
      function of $\nabla T/T$ for the FPU model for $T=8.8$
      (left) and the $\phi^4$ theory for $T=1$ (right) for
      $d=1$.  The dashes represent linear response and the solid
      line show the quadratic deviations from it of the form
      $\kappa \nabla T\left[1+ C_{LR}(\nabla T/T)^2\right]$.
      }
    \label{fig:fourier}
  \end{center}
\end{figure}
When the temperature gradient is small, linear
response theory is applicable so that Fourier's law is
satisfied {\it globally}: $J_0=-\kappa(T)
(T_2-T_1)/L$\cite{ak-long,ak-fpu}.   As the gradient increases,
curvature develops. When $\kappa\sim T^{-\gamma}$, we
can integrate Fourier's law to obtain the next leading order
correction due to curvature in $T(x)$:
\begin{equation}
  \label{lr-breaking}
  {J_{LR}-J_0\over J_0}= \frac{\gamma(\gamma+1)}{24}L^2  
  \left(\frac{\nabla T}{T}\right)^2+\cdots.
\end{equation}
This $L^2$ dependence simply indicates that Fourier's law is
satisfied {\it locally} rather than globally. 
As the gradient increases even further, the energy that can be pumped
through the system becomes less than that predicted by 
linear response theory {\it even when it is applied
  locally}\cite{ak-long}. This is exactly the deviation we study 
here in \figno{fourier}.
This is quite difficult to measure, 
since unlike the cumulants, the theoretical value for the energy
flow $J_{LR}$ is not known and it needs to be obtained using linear
response locally,  which carries an error in itself. 
The relative deviation from the linear response result (as shown
in \figno{fourier}) can be reasonably well explained by
\begin{equation}
  \label{linear-response}
  \delta_{\scriptsize LR}=  {J-J_{LR} \over J_{LR}} = 
  C_{LR}\left(\frac{\nabla T}{T}\right)^2, 
\end{equation}
where $C_{LR}^\phi=-4(3) L ^{1.0(2)}\ (T=1)$ and
$C_{LR}^{FPU}=-6.6(8)L^{0.9(1)}\ (T=8.8)$.  For large gradients,
Eq.~\eqnn{linear-response} will naively give rise to decreasing
current with increasing gradient. However, when the gradient is
this large, the higher order terms in $(\nabla T/T)$ becomes as
important.  We do not know how the system behaves under such an
extremely large gradient, but it would be natural to expect that 
the current will saturate.  
The quadratic behavior in \eqnn{linear-response} is also seen
for both the $\phi^4$ theory and for the FPU~model in $d=1$ at
other temperatures and also for the models in $d>1$, even though 
the extraction of $C_{LR}$ involves larger errors in those cases.
 
Within error, we see that the violation of linear response and
local equilibrium are closely connected, occurring in the same manner:
\begin{equation}
  \delta_{LE}\sim \delta_{LR}.
\end{equation}
Local equilibrium and linear response have no threshold, and
break down at the same rate.

\begin{figure}[htbp]
  \begin{center}
    \leavevmode    \epsfxsize=8cm\epsfbox{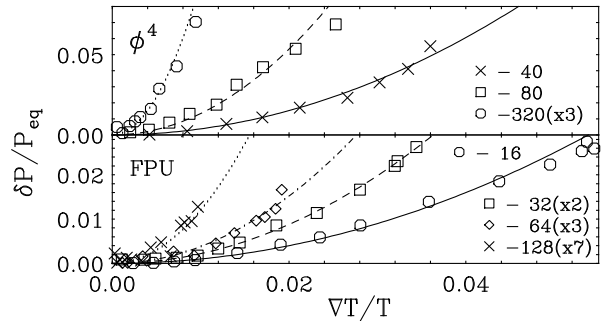}
    \caption{Pressure $P$ as a function of 
      $\nabla T/T$, away from equilibrium for $\phi^4$ (top, $T=1$) and
      FPU (bottom, $T=8.8$). The
      fits of the data to a behavior const.$\times (\nabla
      T/T)^2$ are plotted for each lattice size. $P$ is 
      seen to increase away equilibrium in both
      models. (Vertical axis scaled as indicated.)}
    \label{fig:t11}
  \end{center}
\end{figure}

\noindent\underline{Equation of State:} The equation of state 
for these models are simple in equilibrium since there is only
one independent variable, which can be taken to be $T$. 
(In the FPU model in $d=1$,
there is also a possibly weak dependence on $L$ in $P_{eq}$;
unlike the $L$ dependence of $\kappa$, it is far less
discernible.) We denote it 
$P(T)$ (or $E(T)$) where $P$ ($E$) is the pressure (energy
density). We can measure them through the stress tensor, 
$P=P_{xx}={\cal T}^{11}$,
$E={\cal T}^{00}$\cite{ak-long}.
In \figno{t11}, where we plot the relative deviation of the pressure from its
equilibrium value, $(P-P_{eq})/P_{eq}$, against $\nabla T/T$. We
see that the equation  of state develops new dependences of the form
\begin{equation}
  \label{noneqP}
 P(T,\nabla T,L) = P_{eq}(T)
 \left[ 1+ C_P\left(\frac{\nabla T}{T}\right)^2\right]
\end{equation}
where $C_P^\phi=1.5(1.2)L^{0.9(2)} \ (T=1)$, $C_P^{\scriptstyle
  FPU}=4.1(6)L^{0.30(4)}\ (T=8.8) $. The non-equilibrium
equation of state, $P(T,\nabla T,L)$ develops a non-trivial size
dependence in $C_P$, rendering it non-local. Similar analysis
for energy density yields the coefficients
$C_E^\phi=0.5(3)L^{0.9(2)} \ (T=1)$, $C_E^{\scriptstyle
  FPU}=1.7(7)L^{0.3(1)}\ (T=8.8)$.
The quadratic behavior as in \eqnn{noneqP} is seen also in both
models at different $T$ for $d=1-3$ but we do not have enough
statistics to unambiguously extract $C_{P,E}$ in those cases.
While Extended Irreversible Thermodynamics (EIT) predicts the
quadratic dependence in \eqnn{noneqP} for particle gases and
liquids, precise identification of the non-equilibrium
definitions of the physical quantities between the two theories
is necessary before quantitative comparisons can be made.
EIT predicts a local behavior for $\delta P$
in contrast to our observations.

We have quantified the violations of local equilibrium and linear
response in $\phi^4$ theory and the FPU $\beta$ model in $d=1-3$
dimensions, and observed that they break down at the same rate,
with similar order of magnitude. Both are found to vary with the
thermal gradient as $(\nabla T/T)^2$, and there is no threshold
for the violations, appearing immediately as one moves away from
global equilibrium. Other physical quantities such as the
pressure and the energy density were also found to behave in a
similar manner.  We found that using the coefficients
$C_{\cal A}$, we can predict the spatial dependence of
non-equilibrium distributions of observables. As a consequence,
coarse graining does not modify our conclusions.

Since the definition of $T$ is no longer unique when local
equilibrium is broken, a question arises as to
how the choice of non-equilibrium definition for 
$T$ affects the results.
Expressions for non-equilibrium deviations of $J,P,E$ as in
Eqs.~\eqnn{linear-response},\eqnn{noneqP} will in general be
affected covariantly; in particular, for a generic redefinition
$T=T'+ \nu\left(\nabla T/T\right)^2$, $C'_{\cal A}=C_{\cal
  A}+\nu(dA/dT)$. The local equilibrium violations seen in 
$\cum{p^n}/T^{n/2}$ as in Eq.~\eqnn{le-breaking} are {\it
  invariant} under such redefinitions, up to the order we
consider.  The physics, of course, is invariant under any
redefinition in temperature.
We find that the momentum cumulants provide the most natural and
also the most clear criterion for the violation of local
equilibrium.

Certainly more questions remain: Importantly, we do not have an
analytic understanding of how the coefficients
$C_{LE},C_{LR},C_{P},C_{E}$ are 
related to the parameters of the theories, including their $L$
dependence. 
Another question to consider is how other physical quantities
behave away from local equilibrium.
It would be interesting to further explore the
consequences of non-locality in the equation of state, as well
as use such models as a testing ground for non-equilibrium
thermodynamics, since there is good control on the
non-equilibrium steady state.

\end{document}